\documentclass{JHEP3} 

\usepackage{amsmath,amssymb}
\def\beq{\begin{equation}}
\def\eeq{\end{equation}}
\def\bea{\begin{eqnarray}}
\def\eea{\end{eqnarray}}
\def\nn{\nonumber}
\def\del{\partial}

\title{Lehmann-Symanzik-Zimmermann S-Matrix elements on the Moyal Plane}

\author{A. P. Balachandran \\ Department of Physics, Syracuse University, Syracuse,
NY 13244-1130, USA \\ E-mail: \email{bal@phy.syr.edu}}

\author{Pramod Padmanabhan \\ Department of Physics, Syracuse University,
Syracuse, NY 13244-1130, USA
\\ E-mail: \email{ppadmana@syr.edu}}

\author{Amilcar R. de Queiroz \\ Instituto de Fisica, Universidade de Brasilia, Caixa Postal 04455, 70919-970, Brasilia, DF, Brazil
 \\ E-mail: \email{amilcarq@unb.br}}

\preprint{SU-4252-916}

\abstract
{Field theories on the Groenewold-Moyal(GM) plane are studied using the Lehmann-Symanzik-Zimmermann(LSZ) formalism. The example of real scalar fields is treated in detail. The S-matrix elements in this non-perturbative approach are shown to be equal to the interaction representation S-matrix elements. This is a new non-trivial result: in both cases, the S-operator is independent of the noncommutative deformation parameter $\theta_{\mu\nu}$ and the change in scattering amplitudes due to noncommutativity is just a time delay. This result is verified in two different ways. But the off-shell Green's functions do depend on $\theta_{\mu\nu}$. In the course of this analysis, unitarity of the non-perturbative S-matrix is proved as well.} 
 
\begin{document}
\section{Introduction}

Spacetime at the Planck scale is possibly noncommutative. Physical arguments suggest this possibility~\cite{Sergio}. A noncommutative spacetime which may model such noncommutativity is described by the Moyal algebra~\cite{Sergio}. This noncommutative algebra $\mathcal{A}_{\theta}$ is given by \beq\label{Moyal}{ [ x_{\mu}, x_{\nu} ] = i\theta_{\mu\nu}, ~~ \mu,~\nu = 0,~1,~2,~3,}\eeq $$\theta_{\mu\nu}= -\theta_{\mu\nu} = \textrm{real constants}.$$

The commutation relations Eq.(\ref{Moyal}) are apparently Lorentz non-invariant. However there does exist a twisted action of the Lorentz group on $\mathcal{A}_{\theta}$ which is compatible with its multiplication map. Thus Eq.(\ref{Moyal}) can be made compatible with a twisted action of the Poincar\'{e} group.

Quantum field theories (qft's) on $\mathcal{A}_{\theta}$ have been extensively studied in the past~\cite{Shiraz, BalAnoshAkofor, Rivesseau, Connes}. Different approaches have been used to study them. The initial ones starting from~\cite{Shiraz} were based on the star product approach. There were others using the Seiberg-Witten map~\cite{Seiberg} of the noncommutative theory to a commutative one.
Most of these approaches were plagued by the phenomenon of UV/IR mixing as was first shown in~\cite{Shiraz}. There were also questions regarding the renormalizability of these field theories. The approaches of ~\cite{Rivesseau, Grosse1} restored renormalizability by using a different propagator and interaction for these theories. They also proved that their formulation of scalar field theory is renormalizable to all orders~\cite{Grosse2, Rivasseau1}. In another line of development, with the appearance of the possibility of a twisted action of the Lorentz group on the Moyal plane~\cite{Chaichian}, it was quickly realized by Balachandran and coworkers that the statistics of the quantum fields have to be twisted in order to be compatible with the deformed symmetry group of the noncommutative spacetime~\cite{BalBabar}. As a consequence the twisted perturbative S-matrix was shown to be independent of the noncommutative parameter $\theta_{\mu\nu}$~\cite{BalBabarSasha} in the absence of gauge fields. However when there is an interaction among non-abelian gauge and matter fields, the $\theta$-dependence and UV/IR mixing reappear~\cite{BalBabarGauge2}.

These models are interesting on the phenomenological side as well. As they are theories which violate CPT~\cite{BalCPT} and Lorentz invariance and can lead to Pauli-forbidden transitions~\cite{BalPp, BalAnoPp}, they help us to model the latter as well. They also lead to anisotropies in the CMB spectrum~\cite{Cmb1, Cmb2}. For detailed reviews of the physics on the Moyal plane covering both the theoretical and the phenomenological aspects see~\cite{BalAnoshAkofor, BalPpRev1, BalPpRev2, BalMar}.

Qft's on the Moyal plane can be extended to include gauge fields as well~\cite{BalBabarGauge1}. The gauge fields in the approaches of~\cite{BalBabarGauge1} are not twisted unlike the matter fields and so the gauge group remains the same as in the commutative theory. This circumvents a problem faced in alternative formulations of gauge theories on the Moyal plane where the finite-dimensional Lie algebra of the group of the gauge theory gets enlarged into an infinite dimensional algebra. One important consequence of the twisted field approach of~\cite{BalBabarGauge1} to gauge theories is the addition of a central element to the spacetime symmetry algebra of the system. This results in a new deformed Hopf algebra with a new coproduct. This coproduct does not obey the coassociativity condition. This makes the spacetime also nonassociative~\cite{BalBabarGauge1}.

In the present work, we concentrate on the twisted scalar field theory on the Moyal plane in the absence of gauge fields. We compute the S-matrix elements of this theory using the LSZ reduction formula for the noncommutative case developed in~\cite{BalTrgSv}. It was remarked in~\cite{BalTrgSv} that these amplitudes can be computed using the perturbation theory of Wightman functions~\cite{Steinman} with appropriate modifications. However here we do not use the Wightman function perturbation theory, but instead present two nonperturbative ways of computing the scattering amplitudes. The methods relate the commutative and noncommutative scattering amplitudes. When the in- and out- states are momentum eigenstates, the $\theta$-dependence is in the form of an overall phase multiplying the commutative scattering amplitude. It represents a time delay~\cite{buchholz}. The corresponding $\theta$-dependence via the perturbative interaction representation S-matrix elements appears in the form of the same overall phase so that both approaches are mutually consistent.

 We emphasize that the emergence of this consistency is nontrivial since the systematic formulation of the interaction representation from the Heisenberg representation for the Moyal plane is not easy as we indicate later.

The situation with regard the off-shell Green's functions is different. They do depend on $\theta_{\mu\nu}$ on the Moyal plane as we shall see.

The paper is organized as follows. Section 2 briefly recalls what the noncommutative Moyal algebra ($\mathcal{A}_{\theta}$) is and the notion of twisted quantum fields. In section 3 the LSZ formalism is reviewed for both the $\theta_{\mu\nu}=0$ and the $\theta_{\mu\nu}\neq 0$ cases.

Section 4 shows the two nonperturbative methods of computing the scattering amplitudes in the noncommutative case. In section 5 we conclude with a few important remarks. Directions for further work are also indicated in this section.

\section{Twisted Relativistic Quantum Fields on the Moyal plane $\mathcal{A}_{\theta}$}

The Gronenwold-Moyal or Moyal plane is the algebra $\mathcal A_\theta$ of smooth functions
on $\mathbb{R}^{d+1}$ with a twisted (star) product. It can be
written as~\cite{Chaichian, Drinfel'd1, Aschieri} \beq\label{star}{ f \star g := m_{\theta} (f \otimes g) (x) = m_0
( F_{\theta} f \otimes g) (x)} \eeq where $m_0(f \otimes g)(x)
:= f(x) \cdot g(x)$ stands for the usual pointwise multiplication of
the commutative algebra $\mathcal{A}_0$, \beq\label{twist} {
F_{\theta} = \textrm{exp}\Big(\frac{i}{2}\theta^{\mu \nu}
\partial_{\mu} \otimes \partial_{\nu}\Big),} \eeq is called the
{\it Drinfel'd twist element} and $\theta ^{\mu \nu} = -\theta ^{\nu \mu}
=\textrm{constant}$. P. Watts \cite{Watts} and R. Oeckl
\cite{Oeckl} were the first to observe that the star
product in Eq.(\ref{star}) can be cast using an $F_{\theta}$.

 We next briefly explain the notion of twisted Poincar\'{e} symmetry for the Moyal plane.

The proper orthochronous Poincar\'{e} group $\mathcal{P}_+^{\uparrow}$ acts on mutiparticle states through a coproduct which is a homomorphism from $\mathbb{C}\mathcal{P}_+^{\uparrow}$ to $\mathbb{C}\mathcal{P}_+^{\uparrow}\otimes\mathbb{C}\mathcal{P}_+^{\uparrow}$ where $\mathbb{C}\mathcal{P}_+^{\uparrow}$ is the group algebra of $\mathcal{P}_+^{\uparrow}$~\cite{group}. The factors in the tensor product here act through unitary representations of the Poincar\'{e} group on the single particle Hilbert spaces. On the noncommutative spacetime the coproduct should be compatible with the twisted multiplication map. The work of Aschieri {\it et al.} \cite{Aschieri} and Chaichian {\it
et al.} \cite{Chaichian} based on Drinfel'd's original work
\cite{Drinfel'd1} shows that $h\in{\cal P}_+^\uparrow$ acts on ${\cal
A}_\theta ({\mathbb R}^{d+1})$ compatibly with $m_\theta$ i.e, \beq{m_{\theta}(\Delta_{\theta}(h)f\otimes g) = h\cdot m_{\theta}(f\otimes g),~~f,g\in \mathcal{A}_{\theta}(\mathbb{R}^{d+1})}\eeq if its
coproduct is given by
\begin{equation}
\Delta_\theta (h) = F_\theta^{-1} (h \otimes
h) F_\theta,
\end{equation} where $F_{\theta}=e^{-\frac{i}{2}\hat{P}_{\mu}\otimes\theta_{\mu\nu}\hat{P}_{\nu}}$ and $\hat{P}_{\mu}$ is the generator of translations. It is realized as $-i\del_{\mu}$ on functions.
Thus $\Delta_{\theta}(h)$ is a twisted version of the standard coproduct $\Delta_0(h) = h\otimes h$.

Next we define the notion of twisted statistics on the Moyal plane.

 The action of the twisted coproduct is not compatible with the standard flip or statistics operator defined by $\tau_0$. The operator $\tau_0$ flips two elements of $V\otimes V$ where $V$ is a representation space for $\mathbb{C}\mathcal{P}_+^{\uparrow}$: \beq{\tau_0(f\otimes g)= g\otimes f}\eeq where $f, g \in\mathcal{A}_0$. Now $\tau_{0} F_{\theta} =  F_{\theta}^{-1} \tau_{0}$
so that $\tau_0\ \Delta_\theta(h) \neq\ \Delta_\theta(h)
\tau_0$. This shows that the usual statistics operator is not compatible with
the twisted coproduct. Hence it should be changed in quantum theory. Now the new ``twisted'' statistics operator ~\cite{statistics-uv-ir}
\begin{equation}\label{2.5}
\tau_\theta~\equiv~ F_\theta^{-1}\tau_0  F_\theta, \quad
\tau_\theta^2 = {\bf 1}\otimes {\bf 1}
\end{equation}
does commute with the twisted coproduct,
\begin{equation}
\Delta_\theta (h) = F_\theta^{-1} h \otimes
h~F_\theta.
\end{equation}
Hence $\tau_{\theta}$ is an appropriate twisted flip operator and twisted bosons and fermions are to be defined using the projectors $\frac{1}{2}\left(\mathbb{I}\pm\tau_{\theta}\right)$ respectively.

We now define twisted quantum fields $\phi_{\theta}$ which we will use throughout the rest of this paper. Here for simplicity, we assume that they are scalar fields. They are ``covariant''~\cite{covariance} under the twisted action of the Poincar\'{e} group and incorporate the above
twisted statistics in their creation and annihilation operators. Their star products have the important
self-reproducing property \beq{\phi_{\theta}\star\phi_{\theta}\star\cdots\phi_{\theta} (x) = \left(\phi_0(x)\phi_0(x)\cdots\phi_0(x)\right)_{\theta} }\eeq
where on the right, $\phi_0$'s are first multiplied as ordinary fields and then finally twisted as the subsequent
$\theta$ indicates.

  Consider a free untwisted($\theta_{\mu\nu}=0$) scalar field, $\phi_0$ of mass $m$. It has the mode expansion
\beq\label{freecf}{\phi_0(x)=\int d\mu(p)(a_0(p)~e_p(x) +
a^\dagger_0(p)~e_{-p}(x))}\eeq where $e_p(x)=e^{-ip\cdot x}$,
$p\cdot x=p_0x_0-\vec{p}\cdot\vec{x}$, $d\mu(p)=\frac{d^3p}{(2\pi)^3}\frac{1}{\sqrt{2p_0}}$,
 $p_0=|\sqrt{\vec{p}^2+m^2}|$. The creation and annihilation
operators satisfy the standard commutation
relations, the nonvanishing commutator being \beq{a_0(p)~a^{\dag}_0(q) -
a^{\dag}_0(q)~a_0(p) = (2\pi)^3
\delta^3({\vec{p}}-{\vec{q}}).}\eeq

 The one-particle states are defined as \beq\label{1par}{|\vec{p}\rangle = \sqrt{2E_{\vec{p}}}~a^{\dag}_0(p)~|0\rangle,}\eeq
 with $E_{\vec{p}}=p_0$.  The scalar product between two such states is given by \beq{\langle\vec{p}|\vec{q}\rangle = 2E_{\vec{p}}~(2\pi)^3~\delta^3(\vec{p}-\vec{q}).}\eeq
 The completeness relation for the $1$-particle states is given by \beq{\mathbb{I}_{1-\textrm{particle}} = \int\frac{d^3p}{(2\pi)^3}\frac{1}{2E_{\vec{p}}}~|\vec{p}\rangle\langle\vec{p}|.}\eeq

 The quantum mechanical two-particle bosonic states for $\theta_{\mu\nu}=0$ can be constructed from $\phi_0$ as: \bea\label{untwisted} \langle
 0|\phi_0(x_1)\phi_0(x_2)\sqrt{2E_{\vec{q}}}\sqrt{2E_{\vec{p}}}~a^\dagger_0(q)~a^\dagger_0(p)|0\rangle
& = &(1+\tau_0)(e_{\vec{p}}\otimes e_{\vec{q}})(x_1,x_2)
\nn \\  & \equiv & \langle x_1,x_2|p,q\rangle_{S_0}, \\ |p,q\rangle_{S_0}&=&\sqrt{2E_{\vec{q}}}\sqrt{2E_{\vec{p}}}~a^\dagger_0(q) a^\dagger_0(p) |0\rangle_{S_0},\eea where $\tau_0$ is the commutative flip operator.
 Here the right hand side is symmetric in $x_1$ and $x_2$.

The two-particle states in non-commutative quantum field theory
should obey twisted statistics. Using Eq.(\ref{freecf}) as a
guide, we can construct the twisted scalar quantum field
$\phi_{\theta}(x)$ as \beq{\phi_{\theta}(x)=\int
d\mu(p)(a_{\theta}(p)~e_p(x) + a^\dagger_{\theta}(p)~e_{-p}(x))}\eeq
It is possible to write the twisted creation and annihilation
operators $a^\dagger_{\theta}(p), a_{\theta}(p) $ in terms of the
untwisted operators in Eq.(\ref{freecf}). The transformation
connecting the twisted and untwisted creation and annihilation
operators is called the ``dressing transformation" \cite{dressing1,
dressing2} and is given by
\beq{a_{\theta}(p)=a_0(p)~e^{-\frac{i}{2}p_{\mu}\theta^{\mu\nu}P_{\nu}}.}\eeq

Using the above twisted field, we can construct twisted two-particle states
as in Eq.(\ref{untwisted}):  \bea\label{twisted} \langle
0|\phi_{\theta}(x_1)\phi_{\theta}(x_2)\sqrt{2E_{\vec{q}}}\sqrt{2E_{\vec{p}}}~a^\dagger_{\theta}(q)~a^\dagger_{\theta}(p)|0\rangle
& = &(1+\tau_{\theta})(e_{\vec{p}}\otimes e_{\vec{q}})(x_1,x_2)
\nn \\ & \equiv & \langle x_1,x_2|p,q\rangle_{S_{\theta}}, \\ |p,q\rangle_{S_\theta}&=&\sqrt{2E_{\vec{q}}}\sqrt{2E_{\vec{p}}}~a^\dagger_\theta(q) a^\dagger_\theta(p) |0\rangle_{S_\theta},\eea where $\tau_{\theta}$ is the twisted flip operator given in Eq.(\ref{2.5}). Note that the reversed ordering of $p,q$ as we go from LHS to RHS really matters here \cite{BPQ}.
From Eq.(\ref{twisted}) we can deduce the relations~\cite{statistics-uv-ir, covariance} \begin{eqnarray} & a^\dagger_{\theta}(p)~a^{\dag}_{\theta}(q)=
e^{ip_{\mu}\theta^{\mu\nu}q_{\nu}}a^{\dag}_{\theta}(q)~
a^\dagger_{\theta}(p), \\ & a_{\theta}(p)~ a_{\theta}(q)=
e^{ip_{\mu}\theta^{\mu\nu}q_{\nu}}a_{\theta}(q)~ a_{\theta}(p)\end{eqnarray}

Here $P_{\mu}$ is the four-momentum operator:  \beq{P_{\mu}=\int
\frac{d^3p}{(2\pi)^3}(a^\dagger_0(p) ~a_0(p))p_{\mu}~ =\int
\frac{d^3p}{(2\pi)^3}(a^\dagger_{\theta}(p)~ a_{\theta}(p))p_{\mu}. }\eeq
Note that both the twisted and untwisted 4-momentum operators are the same
since $p_{\mu}\theta^{\mu\nu}P_{\nu}$ commutes with $a^\dagger_0(p)~
a_0(p)$.

We can write the twisted quantum field in terms of the untwisted one
with the help of the dressing transformation as
\beq{\phi_{\theta}(x)=\phi_0(x)e^{\frac{1}{2}\overleftarrow{\partial_{\mu}}\theta^{\mu\nu}P_{\nu}}}\eeq

\section{The Untwisted and Twisted LSZ Reduction Formula}

The LSZ formalism for computing scattering amplitudes is non-perturbative. There are two ways to arrive at the formula for scattering amplitudes~\cite{Peskin, Weinberg, Srednicki}. We use the approach given in~\cite{Srednicki}. After discussing it briefly for $\theta_{\mu\nu}=0$, we recall~\cite{BalTrgSv}, where the twisted LSZ reduction formula was derived.

\subsection*{The $\theta_{\mu\nu}=0$ case}

Consider an interacting quantum field theory whose Hamiltonian $H$ can be split as \beq{H = H_0+H_I}\eeq where $H_0$ is the free Hamiltonian for a massive field and $H_I$ is the interaction part. $H_0$ is used to define the states in the infinite past and infinite future. The in- and out-states of the theory are eigenstates of the full Hamiltonian $H$, which evolve like free states in the infinite past and future. On the other hand, free states are eigenstates of the free Hamiltonian $H_0$, whose evolutions are governed by $H_0$ itself.
The LSZ formalism works with the in- and out-states. There are creation-annihilation operators $a_0^{\dag\textrm{in(out)}}(k)$, $a_0^{\textrm{in(out)}}(k)$ which create the in- and out- states. Note that these are not the free creation-annihilation operators. They are used in the mode expansion of the in- and out-fields. They help create the in(out) states $|k_1,k_2,\cdots ,k_N;~\textrm{in(out)}\rangle$.

The interacting vacuum is unique after a phase choice.

The LSZ reduction formula for $\theta_{\mu\nu}=0$ can be now written as
\beq{\langle k_N', ... , k_1' ;~ \textrm{out} | k_M, ... , k_1;~\textrm{in} \rangle = \int \mathcal{I}~G^0_{N+M} (x_1', ... , x_N';~ x_1, ... , x_M),} \eeq
where \beq\label{integral}{\mathcal{I}=\prod_{i=1}^N d^{4}x_i' \prod_{j=1}^M d^{4}x_j~e^{-i(k_j\cdot x_j-k_i'\cdot x_i')}~i(\partial_i'^2+m^2)~i(\partial_j^2+m^2)}\eeq
and \beq{G^0_{N+M}(x_1',..., x_N';~ x_1, \cdots , x_M) = \langle \Omega|T\left[\phi_0 (x_1')\cdots \phi_0 (x_N')\phi_0 (x_1)\cdots \phi_0 (x_M)\right] |\Omega\rangle}\eeq where $|\Omega\rangle$ is the interacting vacuum and $G^0_{N+M}(x_1', ... ,x_N';~ x_1, \cdots ,x_M)$ is the Green's function for $M$ in-fields and $N$ out-fields.
The proof is standard and can be found in textbooks like~\cite{Srednicki}.

Now we write down the twisted LSZ formula.

\subsection*{The $\theta_{\mu\nu}\neq 0$ case}

It was argued in~\cite{BalTrgSv} that the relations between the twisted in- and out-creation-annihilation operators and the free creation-annihilation operators are: \beq{a_{\theta}^{\textrm{in, out}}(k) = a_0^{\textrm{in, out}}(k) e^{-\frac{i}{2}k_{\mu}\theta^{\mu\nu}\hat{P}_{\nu}},}\eeq \beq{a_{\theta}^{\dag\textrm{in, out}}(k) = a_0^{\dag\textrm{in, out}}(k) e^{\frac{i}{2}k_{\mu}\theta^{\mu\nu}\hat{P}_{\nu}}.}\eeq Thus as remarked above, the in- and out-fields can be obtained from the commutative ones from the formula \beq{\phi_{\theta}^{\textrm{in, out}} = \phi^{\textrm{in, out}}_0 e^{\frac{1}{2}\overleftarrow{\partial}_\mu \theta^{\mu \nu} \hat{P}_\nu}.}\eeq   The twisted in- and out-states are created using the twisted in- and out creation-annihilation operators. The twisted LSZ reduction formula is given by~\cite{BalTrgSv} \begin{equation}\label{twistedLSZ} _{\theta}\langle k_N', ... , k_1' ;~ \textrm{out} | k_M, ... , k_1;~\textrm{in} \rangle_\theta = \int \mathcal{I}~G^{\theta}_{N+M} (x_1', ... , x_N';~ x_1, ... , x_M), \end{equation} where  $\mathcal{I}$ is defined in Eq.(\ref{integral}),  and \begin{multline}G_{N+M}^{\theta}(x_1', \cdots , x_N'~;~x_1,\cdots , x_M) = T\left[ e^{-\frac{i}{2}\left[\sum_{i<j}\partial_{z_{i,\mu}}\theta^{\mu \nu}\partial_{z_{j,\nu}} \right]}\times \right . \\ \left . W_{N+M}^0(z_1, \cdots , z_N~;~ z_{N+1}, \cdots , z_{N+M})\right]\end{multline} with \beq\label{zs}{z_i = x_i',~ i\leq N;~~ z_{N+i}=x_i,~i\leq M.}\eeq  In the above $W^0_{N+M}(z_1, \cdots , z_N~;~ z_{N+1}, \cdots , z_{N+M})$ is the Wightman function for $\theta_{\mu\nu}=0$ given by \beq{W^0_{N+M}(z_1, \cdots , z_N~;~ z_{N+1}, \cdots , z_{N+M}) = \langle\Omega |\phi_0(z_1)\cdots \phi_0(z_{N+M})|\Omega\rangle }\eeq where $|\Omega\rangle$ is the exact vacuum of the fully interacting theory, the arguments of the fields are given in Eq.(\ref{zs}) and $\phi_0$'s are the fully interacting commutative quantum fields.

We will use this formula to evaluate scattering amplitudes in the noncommutative case.

\section{Non-perturbative Computations of the Scattering Amplitudes}

In this section, in order to avoid index cluttering, we use notations such as
\begin{equation}
      p_i \wedge p_j \equiv p_{i,\mu}~\theta^{\mu \nu}~p_{j,\nu}, ~~ \partial\wedge P = \partial_{\mu}\theta^{\mu\nu}P_{\nu}
\end{equation}
where $i,j$ stand for particle labels, and $\mu,\nu$ as usual stand for spacetime components.

\subsection*{Method 1}

The in- and out- states for the twisted case are
\begin{eqnarray}\label{in}|p_M,...,p_1;~\textrm{in}\rangle_\theta&=& \sqrt{(2E_{\vec{p}_1})\cdots (2E_{\vec{p}_M})}~a_{\theta}^{\dag\textrm{in}}(p_1) \cdots a_{\theta}^{\dag\textrm{in}}(p_M)|\Omega\rangle  \nn \\ &=& \sqrt{(2E_{\vec{p}_1})\cdots (2E_{\vec{p}_M})}~a_0^{\dag\textrm{in}}(p_1)\cdots a_0^{\dag\textrm{in}}(p_M)|\Omega\rangle e^{\frac{i}{2}\sum_{i<j\leq M}p_i \wedge p_j}\end{eqnarray}
and
\begin{eqnarray}\label{out}|p_1',...,p_N';~\textrm{out}\rangle_\theta&=&\sqrt{(2E_{\vec{p}_1'})\cdots (2E_{\vec{p}_N'})}~a_{\theta}^{\dag\textrm{out}}(p_N') \cdots a_{\theta}^{\dag\textrm{out}}(p_1')|\Omega\rangle  \\ &=& \nn \sqrt{(2E_{\vec{p}_1'})\cdots (2E_{\vec{p}_N'})}~a_0^{\dag\textrm{out}}(p_N') \cdots a_0^{\dag\textrm{out}}(p_1')|\Omega\rangle e^{\frac{i}{2}\sum_{i<j\leq N}p_i'\wedge p_j'}\end{eqnarray}
respectively.

It can now be immediately seen that the twisted scattering amplitude in terms of the untwisted scattering amplitude can be obtained by using the definition of the LSZ $S$-matrix:
\beq{S_{\theta}(p_N',..., p_1';~ p_M,..., p_1) =~ _{\theta}\langle p_N', ... , p_1';~\textrm{out}|p_M, ... , p_1 ;~\textrm{in}\rangle_{\theta}.}\eeq
By using the definition of the twisted in- and out- states given by Eq.(\ref{in}) and Eq.(\ref{out}) respectively, we see that
\begin{eqnarray}_{\theta}\langle p_N', ... , p_1';~\textrm{out}|p_M, ... , p_1 ; \textrm{in}\rangle_{\theta} &=&  e^{\frac{i}{2}\left[\sum_{i<j\leq M} p_i\wedge p_j - \sum_{i<j\leq N}p_i'\wedge p_j'\right]} \times \\ \nn & & _0\langle p_N', ... , p_1';~\textrm{out}|p_M, ... , p_1 ; \textrm{in}\rangle_0.\end{eqnarray}
Thus the twisted scattering amplitude for any process is given by
\begin{eqnarray}\label{shortest}S_{\theta}(p_N',..., p_1';~ p_M,..., p_1) &=&  e^{\frac{i}{2}\left[\sum_{i<j\leq M} p_i\wedge p_j - \sum_{ i<j\leq N}p_i'\wedge p_j'\right]}\times \\ \nn & & S_0(p_N',..., p_1';~ p_M,..., p_1).\end{eqnarray}
This relation between the commutative and the noncommutative scattering amplitudes is the same as the one obtained via the interaction representation formalism \cite{BalBabarSasha,Amilcar}.

We note that this method is non-perturbative and is completely independent of the interaction term in the scalar field theory considered.

The scattering amplitude on the Moyal plane given by Eq.(\ref{shortest}) also shows that the twisted $S$-matrix is unitary in a trivial way, since the commutative $S$-matrix is unitary.

\subsection*{Method 2}

In this second method we will find the same result via the reduction formula. It brings out the difference between scattering amplitudes and off-shell Green's functions.

The computation shown here closely follows the derivation of the reduction formula given in~\cite{Peskin}.

Here we will consider as an example the time ordered product of four fields representing a process of two particles going into two other particles described by the correlation function \beq{G_{2+2}^0(x_1', x_2';~x_1,x_2)=\langle\Omega|T\left(\phi_0(x_1')\phi_0(x_2')\phi_0(x_1)\phi_0(x_2)\right)|\Omega\rangle}\eeq which is the appropriate Green's function for the untwisted case. The Green's functions for the twisted case is obtained by replacing the commutative fields by the noncommutative ones and $G_{2+2}^0$ by $G^{\theta}_{2+2}.$ The procedure involves finding the pole structure in momentum space of the Fourier transform of $G_{2+2}^0(x_1', x_2';~x_1, x_2)$.

We first consider the commutative case.

\subsubsection*{$\theta_{\mu\nu}=0$}

 Let us consider the general off-shell Fourier transforms
\beq\label{fourier}\int \prod_{i=1}^j d^4x_i'~e^{ip_i'\cdot x_i'}G_{N+M}^0(x_1',..., x_N';~x_1,..., x_M) = \\ \tilde{G_0}^{(j)}(p_1',... ,p_j',... , x_N', x_1,...., x_M).\eeq

 Consider Fourier transforming $G_{2+2}^0(x_1', x_2';~x_1, x_2)$ in just $x_1'$. Assume without loss of generality that $x_1'^0$ is associated with an outgoing particle. Split the $x_1'^0$-integral into three regions as follows:
\beq{\left( \int_{T_+}^{\infty}dx_1'^0 + \int_{-\infty}^{T_-}dx_1'^0 + \int_{T_-}^{T_+}dx_1'^0 \right)d^3x_1'~e^{ip_1'^0x_1'^0 - i \vec{p}_1'\cdot\vec{x}_1'} ~G_{2+2}^0(x_1', x_2';~x_1, x_2).}\eeq
Here $T_+>> \textrm{max}(x_2'^0, x_1^0, x_2^0)$ and $T_-<<\textrm{min}(x_2'^0, x_1^0, x_2^0)$.
Since $T_+\geq x_1'^0 \geq T_-$ is a finite interval, the corresponding integral will not give any pole.
A pole comes from single particle insertion in the integral over $x_1'^0\geq T_+$ in $G_{2+2}^0$ as we now show following~\cite{Peskin}.  In the integration between the limits $T_+$ and $+\infty$, $\phi(x_1')$ stands to the extreme left inside the time-ordering so that
\beq{G_{2+2}^0(x_1', x_2';~x_1, x_2)= \int\frac{d^3q_1}{(2\pi)^3}\frac{1}{2E_{\vec{q}_1}} \langle\Omega|\phi_0(x_1')|q_1\rangle\langle q_1|T\left(\phi_0(x_2')\phi_0(x_1)\phi_0(x_2)\right)|\Omega\rangle + \textrm{OT}}\eeq
where $\textrm{OT}$ stands for the other terms. These other terms include those which arise from the omitted time orderings.

The matrix element of the field $\phi_0(x_1')$ can be written as
\begin{eqnarray} \langle\Omega|e^{iP\cdot x_1'}\phi_0(0)e^{-iP\cdot x_1'}|E_{\vec{q}_1}, \vec{q_1}\rangle & = \langle\Omega|\phi_0(0)|E_{\vec{q_1}}, \vec{q_1}\rangle e^{-iq_1\cdot x_1'}|_{q_1^0=E_{\vec{q}_1}}  \nn \\ & = \langle\Omega|\phi_0(0)|q^0_1,\vec{q}_1=0\rangle e^{-iq_1\cdot x_1'}|_{q_1^0=E_{\vec{q}_1}}\end{eqnarray} where $E_{\vec{q_1}}= \sqrt{\vec{q_1}^2+m^2}$.  In obtaining the above relation we have used the Lorentz invariance of the vacuum and of $\phi_0(0)$~\cite{Peskin}.
Thus
\beq{\langle\Omega|\phi_0(x_1')|E_{\vec{q_1}}, \vec{q}_1\rangle = \sqrt{Z} e^{-i\left(E_{\vec{q_1}}x_1'^0-\vec{q}_1\cdot\vec{x}_1'\right)}}\eeq
where
\beq{\langle\Omega|\phi_0(0)|q^0_1,\vec{q}_1=0\rangle = \sqrt{Z}}\eeq
and $q_1^0 > 0$. In the above $\sqrt{Z}$ is the field-strength renormalization factor.  So the integral between $T_+$ and $+\infty$ becomes
\beq{\sqrt{Z}\frac{1}{2E_{\vec{p'}_1}}\int_{T_+}^{\infty}dx_1'^0 ~e^{i\left(p_1'^0 - E_{\vec{p}_1'} +i\epsilon\right)x_1'^0}~\langle p_1'|T\left(\phi_{2'}\phi_1\phi_2\right)|\Omega\rangle + \textrm{OT}}\eeq
where $\epsilon > 0$ is the adiabatic cut-off and $\phi_0(x_i)=\phi_i$. Performing the $x_1'^0$ integral we get
\beq{\tilde{G_0}^{(1)}(p_1', x_2', x_1, x_2)= \sqrt{Z}\frac{i}{2E_{\vec{p'}_1}} \frac{e^{i\left(p_1'^0-E_{\vec{p}_1'}+i\epsilon\right)T_+}}{\left(p_1'^0-E_{\vec{p}_1'}+i\epsilon\right)}\langle p_1'|T\left(\phi_{2'}\phi_1\phi_2\right)|\Omega\rangle + \textrm{OT}}\eeq
which as $p_1'^0 \rightarrow E_{\vec{p}_1'}$, becomes
\beq\label{pole}{\tilde{G_0}^{(1)}(p_1', x_2', x_1, x_2) = \sqrt{Z}  \frac{i}{p_1'^2-m^2-i\epsilon}~\langle p_1'|T\left(\phi_{2'}\phi_1\phi_2\right)|\Omega\rangle + \textrm{OT}.}\eeq

In the integration over $(-\infty, T_-)$, $\phi_0(x_1')$ will stand to the extreme right in the time ordered product, so the one-particle state contribution comes from
\beq{\langle q_1|\phi_0(x_1')|\Omega\rangle = \sqrt{Z}e^{i\left(E_{\vec{q}_1}x_1'^0-\vec{q}_1\cdot\vec{x}_1'\right)}.}\eeq
The energy denominator is thus $\frac{1}{p_1'^0+E_{\vec{p}_1'}-i\epsilon}$ and has no pole for $p_1'^0>0$. Thus the answer for the pole is given by Eq.(\ref{pole}).

For the two-particle scattering $p_1, p_2\rightarrow p_1', p_2'$, we can now proceed similarly. The poles appear in both $p_1'^0$ and $p_2'^0$ when both $x_1'^0$ and $x_2'^0$ integrations are large:
\beq{ x_1'^0,~ x_2'^0~ >>~ T_1~>>~x_1^0,~ x_2^0.}\eeq
So for these poles
\begin{multline} \tilde{G_0}^{(2)}(p_1', p_2', x_1, x_2) = \int_{T_+}^{\infty} dx_1'^0 dx_2'^0 d^3x_1' d^3x_2'~ e^{ip_1'\cdot x_1' + ip_2'\cdot x_2'}\frac{1}{2!}\left(\frac{1}{(2\pi)^3}\right)^2\frac{d^3q_1 d^3q_2}{(2E_{\vec{q}_1})(2E_{\vec{q}_2})}\times \\ \langle\Omega|\phi_0(x_1')\phi_0(x_2')|\vec{q}_1\vec{q}_2\rangle\langle\vec{q}_1\vec{q}_2|T\left(\phi_1\phi_2\right)|\Omega\rangle + \textrm{OT}.\end{multline}
Here $T_+$ is considered to be very large. We set $\phi_0(x_1')$, $\phi_0(x_2')$ to be out fields. As we set $|\vec{q}_2\vec{q}_1\rangle$ to $|\vec{q}_2\vec{q}_1\rangle_{\textrm{out}}$ for large $T_+$, only $\langle\Omega|\phi^{\textrm{out}+}_0(x_1')\phi^{\textrm{out}+}_0(x_2')|\vec{q}_2\vec{q}_1\rangle_{\textrm{out}}$, where $\phi_0^{\textrm{out}+}$ is the annihilation part of the out-field, contributes. Thus there is no time-ordering needed involving these out-fields. So we have
\begin{multline}\tilde{G_0}^{(2)}(p_1',p_2',x_1,x_2) = \int_{T_+}^{\infty}d^4x_1'd^4x_2' ~e^{ip_1'\cdot x_1'+ip_2'\cdot x_2'}\frac{1}{2!}\left(\frac{1}{(2\pi)^3}\right)^2\left(\frac{d^3q_1}{2E_{\vec{q}_1}}\right)\left(\frac{d^3q_2}{2E_{\vec{q}_2}}\right)\times \\ \langle\Omega|\phi_0^{\textrm{out}}(x_1')\phi_0^{\textrm{out}}(x_2')|\vec{q}_2\vec{q}_1\rangle_{\textrm{out}}~_{\textrm{out}}\langle\vec{q}_2\vec{q}_1|T\left(\phi_1\phi_2\right)|\Omega\rangle.\end{multline}
Now
\beq{\langle\Omega|\phi_0^{\textrm{out}}(x_1')\phi_0^{\textrm{out}}(x_2')|\vec{q}_2\vec{q}_1\rangle_{\textrm{out}}=\langle\Omega|\phi_0^{\textrm{out}}(x_1')|\vec{q}_1\rangle\langle\Omega|\phi_0^{\textrm{out}}(x_2')|\vec{q}_2\rangle + \vec{q}_2\leftrightarrow\vec{q}_1.}\eeq
Thus Eq.(\ref{pole}) generalizes to
\begin{multline}\tilde{G_0}^{(2)}(p_1',p_2', x_1,x_2)=\left[\sqrt{Z} \left(\frac{i}{p^{'2}_1-m^2-i\epsilon}\right)\right]\left[\sqrt{Z} \left(\frac{i}{p^{'2}_2-m^2-i\epsilon}\right)\right]\times \\ _{\textrm{out}}\langle\vec{p}_1' \vec{p}_2'|T\left(\phi_1\phi_2\right)|\Omega\rangle + \textrm{OT}.\end{multline}

With similar calculations for incoming poles, with $x_1^0, x_2^0 << T_- << x_1'^0, x_2'^0$,
\begin{multline}\label{commutativeLSZ}\tilde{G_0}^{(4)}(p_1', p_2', p_1, p_2) = \prod_{i=1}^2\prod_{j=1}^2 \left[\sqrt{Z}  \left(\frac{1}{p_i^{'2}-m^2-i\epsilon}\right)\right]\left[\sqrt{Z} \left(\frac{1}{p_j^2-m^2-i\epsilon}\right)\right]\times \\ _{\textrm{out}}\langle p_1'~ p_2'~|~ p_1~p_2\rangle_{\textrm{in}}\end{multline}
as required.

\subsubsection*{$\theta_{\mu\nu}\neq 0$}

We will work along lines similar to the one followed for the commutative case to arrive at the twisted version of Eq.(\ref{commutativeLSZ}). However the process we consider in the noncommutative case will not be a 2-particle scattering process as chosen in the commutative case.
Instead we consider a process where $M$ particles go into $N$ particles. 

We introduce the following notations: \beq{ \hat{p}~ \textrm{is an on-shell momentum}= (E_{\vec{p}}=\sqrt{\vec{p}^{~2}+m^2},~ \vec{p})}\eeq
\beq{p~ \textrm{is a generic 4-momentum, with}~ p^0>0.}\eeq

\subsubsection*{Completeness}

The completeness relations for the twisted in- and out-states are the same as in the commutative case, since the noncommutative phases cancel each other. Hence \begin{multline}\label{5.25} a_{\theta}^{\dag\textrm{in, out}}(p_N)\cdots a_{\theta}^{\dag\textrm{in, out}}(p_1)|\Omega\rangle\langle\Omega|a^{\textrm{in, out}}_{\theta}(p_1)\cdots a^{\textrm{in, out}}_{\theta}(p_N) = \\  a_0^{\dag\textrm{in, out}}(p_N)\cdots a_0^{\dag\textrm{in, out}}(p_1)|\Omega\rangle\langle\Omega|a^{\textrm{in, out}}_0(p_1)\cdots a^{\textrm{in, out}}_0(p_N).\end{multline} From Eq.(\ref{5.25}) follow both the resolution of identity given below and hence completeness for the twisted in- and out-states.

\subsubsection*{Resolution of Identity}

 Consider \beq{I' = \sum_N\frac{1}{N!}\left(\int \prod_{i=1}^N \frac{d^3p_i }{(2\pi)^3}\frac{1}{2E_{\vec{p_i}}}\right) a_{\theta}^{\dag\textrm{in, out}}(p_N)\cdots a_{\theta}^{\dag\textrm{in, out}}(p_1)|\Omega\rangle\langle\Omega|a^{\textrm{in, out}}_{\theta}(p_1)\cdots a^{\textrm{in, out}}_{\theta}(p_N).}\eeq This is independent of $\theta_{\mu\nu}$ due to Eq.(\ref{5.25}) and hence is the resolution of identity: \beq\label{id}{I'=I = \sum_N\frac{1}{N!}\left(\int \prod_{i=1}^N \frac{d^3p_i}{(2\pi)^3}\frac{1}{2E_{\vec{p_i}}}\right) a_0^{\dag\textrm{in, out}}(p_N)\cdots a_0^{\dag\textrm{in, out}}(p_1)|\Omega\rangle\langle\Omega|a^{\textrm{in, out}}_0(p_1)\cdots a^{\textrm{in, out}}_0(p_N).}\eeq

For the scattering process of $M$ particles to $N$ particles, the twisted $N+M$-point Green's function we need to look at is
\beq{G^{\theta}_{N+M}(x_1',..., x_N';~x_1,..., x_M) = \langle\Omega|T\left(\phi_{\theta}(x_1')\cdots\phi_{\theta}(x_N')\phi_{\theta}(x_1)\cdots\phi_{\theta}(x_M)\right)|\Omega\rangle.}\eeq
This is Fourier transformed by integrating with respect to the measure
$$ \left(\prod_i d^4x_i'\right)\left(\prod_j d^4x_j\right) e^{i\left(\sum_{i\leq N}p_i'\cdot x_i' - \sum_{j\leq M}p_j\cdot x_j\right)}. $$

Integration over $x_i$, $x_i'$ gives $\tilde{G_{\theta}}^{N+M}(p_1'\cdots , p_N', p_1 \cdots ,p_M)$ and the residue at the poles in all the momenta multiplied together gives the scattering amplitude. This is just the noncommutative version of the LSZ reduction formula. We show that we obtain the same answer as Method I for the S-matrix elements in this way.

\subsubsection*{Pole in just $p_1'$}

Fourier transform just in $x_1'$ to obtain
\begin{multline}\tilde{G_{\theta}}^{(1)}(p_1',\cdots ,x_N', x_1, \cdots ,x_M) = \int d^4x_1' ~e^{i\left(p_1'^0x_1'^0 - \vec{p}_1'\cdot\vec{x}_1'\right)}\times \\ \langle\Omega|T\left(\phi_{\theta}(x_1')\cdots\phi_{\theta}(x_N')\phi_{\theta}(x_1)\cdots\phi_{\theta}(x_M)\right)|\Omega\rangle.\end{multline}
With $T_+ >> x_N'^0\cdots x_2'^0, x_M^0, \cdots ,x_1^0$, we isolate the term with pole in $\tilde{G_{\theta}}^{(1)}$:
\begin{multline} \tilde{G_{\theta}}^{(1)}(p_1',\cdots ,x_N', x_1\cdots x_M) = \sqrt{Z}\int_{T_+}^{\infty} dx_1'^0 d^3x_1'~ e^{i\left(p_1'^0x_1'^0 - \vec{p}_1'\cdot\vec{x}_1'\right)}\times \\ \langle\Omega|\phi_{\theta}^{\textrm{out}}(x_1')T\left(\phi_{\theta}(x_2')\cdots\phi_{\theta}(x_N')\phi_{\theta}(x_1)\cdots\phi_{\theta}(x_M)\right)|\Omega\rangle + \textrm{OT} \\ = \sqrt{Z}\int_{T_+}^{\infty}dx_1'^0d^3x_1'\frac{1}{(2\pi)^3} \frac{d^3q_1}{2E_{\vec{q}_1}}e^{i\left(p_1'^0x_1'^0 - \vec{p}_1'\cdot\vec{x}_1'\right)}\times \\  \langle\Omega|\phi_{\theta}^{\textrm{out}}(x_1')|\hat{q}_1\rangle\langle\hat{q}_1|T\left(\phi_{\theta}(x_2')\cdots\phi_{\theta}(x_N')\phi_{\theta}(x_1)\cdots\phi_{\theta}(x_M)\right)|\Omega\rangle + \textrm{OT} \end{multline}
where
\beq{\langle\Omega|\phi_{\theta}^{\textrm{out}}(x_1')|\hat{q}_1\rangle = \langle\Omega|\phi_0^{\textrm{out}}(x_1')|\hat{q}_1\rangle }\eeq
as the twist gives just 1 in this case. This can be seen by writing $\phi_{\theta}^{\textrm{out}}$ as $e^{\frac{1}{2}\partial_{\mu}\theta^{\mu\nu}P_{\nu}}\phi_{\theta}^{\textrm{out}}$ and acting with $P_{\nu}$ on $\langle\Omega|$. 


Repeating the same procedure as in that of the commutative case, we can extract the pole $\frac{1}{p_1'^2-m^2-i\epsilon}$ and its coefficient.

\subsubsection*{Extracting poles at $p_1'$, $p_2'$}

 In this case we are led to
 \begin{multline}\tilde{G_{\theta}}^{(2)}(p_1', p_2', x_3', \cdots , x_N', x_1, \cdots ,x_M) = \int_{T_+}^{\infty} d^4x_1' d^4x_2'~ e^{ip_1'\cdot x_1' + ip_2'\cdot x_2'} (\sqrt{Z})^2\frac{d^3\hat{q}_1d^3\hat{q}_2}{2! (2E_{\vec{q}_1})(2E_{\vec{q}_2})}\times \\  \langle\Omega|\phi_{\theta}^{\textrm{out}}(x_1')\phi_{\theta}^{\textrm{out}}(x_2')|\hat{q}_1, \hat{q}_2\rangle\langle\hat{q}_1, \hat{q}_2|T\left(\phi_{\theta}(x_3')\cdots\phi_{\theta}(x_N')\phi_{\theta}(x_1)\cdots\phi_{\theta}(x_M)\right)|\Omega\rangle  + \textrm{OT}.\end{multline} Note that there is no twist in $|\hat{q}_1, \hat{q}_2\rangle$ and $\langle\hat{q}_2, \hat{q}_1|$ (See Eq.(\ref{id})).

  We now compute the matrix element of the two out-fields.
  \begin{multline} \langle\Omega|\phi_{\theta}^{\textrm{out}}(x_1')\phi_{\theta}^{\textrm{out}}(x_2')|\hat{q}_1, \hat{q}_2\rangle = \int \left(\frac{1}{(2\pi)^3}\right)^2 \frac{d^3p_1''}{\sqrt{2E_{\vec{p''}_1}}}\frac{d^3p_2''}{\sqrt{2E_{\vec{p''}_2}}}e^{-i\hat{p}_1''\cdot x_1' - i\hat{p}_2''\cdot x_2'} \times \\ \sqrt{2E_{\vec{q}_1}}\sqrt{2E_{\vec{q}_2}}\langle\Omega|\left(a_0^{\textrm{out}}(p_1'')e^{-\frac{i}{2}\hat{p}_1''\wedge P}\right)\left(a_0^{\textrm{out}}(p_2'')e^{-\frac{i}{2}\hat{p}_2''\wedge P}\right) a^{\dag\textrm{out}}_0(\hat{q}_2)a^{\dag\textrm{out}}_0(\hat{q}_1)|\Omega\rangle \\ = \int \left(\frac{1}{(2\pi)^3}\right)^2 \frac{d^3p_1''}{\sqrt{2E_{\vec{p''}_1}}}\frac{d^3p_2''}{\sqrt{2E_{\vec{p''}_2}}}e^{-i\hat{p}_1''\cdot x_1' - i\hat{p}_2''\cdot x_2'} e^{-\frac{i}{2}\hat{p}_1''\wedge\left(-\hat{p}_2''+\hat{q}_1 + \hat{q}_2\right)} e^{-\frac{i}{2}\hat{p}_2''\wedge\left(\hat{q}_1 + \hat{q}_2\right)}\times \\ \sqrt{2E_{\vec{q}_1}}\sqrt{2E_{\vec{q}_2}}\langle\Omega|a_0^{\textrm{out}}(p_1'')a_0^{\textrm{out}}(p_2'')a^{\dag\textrm{out}}_0(q_2)a^{\dag\textrm{out}}_0(q_1)|\Omega\rangle. \end{multline}
  The matrix element becomes
  \begin{multline}\langle\Omega|a_0^{\textrm{out}}(p_1'')a_0^{\textrm{out}}(p_2'')a^{\dag\textrm{out}}_0(q_2)a^{\dag\textrm{out}}_0(q_1)|\Omega\rangle = \left(2\pi\right)^3\left(2\pi\right)^3 \left[\delta^3(\vec{p''}_1-\vec{q}_1)\delta^3(\vec{p''}_2-\vec{q}_2)  \right. \\ \left. + \delta^3(\vec{p''}_1-\vec{q}_2)\delta^3(\vec{p''}_2-\vec{q}_1)\right]\end{multline}
 which means that the whole matrix element is 0 unless
 \beq{\hat{p}_1'' + \hat{p}_2'' = \hat{q}_1 + \hat{q}_2.}\eeq
 So the noncommutative phase can be simplified according to 
 \beq{e^{-\frac{i}{2}\hat{p}_1''\wedge\left(-\hat{p}_2'' + \hat{p}_1'' + \hat{p}_2''\right) -\frac{i}{2}\hat{p}_2''\wedge\left(\hat{p}_1'' + \hat{p}_2''\right)} = e^{-\frac{i}{2}\hat{p}_2'' \wedge\hat{p}_1''}.}\eeq
 Integrations over $\vec{x}_1'$, $\vec{x}_2'$ give $\delta$-functions setting
 \beq{\vec{p''}_1=\vec{p'}_1~,~ \vec{p''}_2 = \vec{p'}_2}\eeq
 and hence
 \beq{\hat{p}_1''=\hat{p}_1'~,~ \hat{p}_2'' = \hat{p}_2'.}\eeq
Thus the noncommutative phase becomes $e^{-\frac{i}{2}\hat{p}_2'\wedge\hat{p}_1'}$.

 Since
 \beq{_{\textrm{out}}\langle\hat{q}_1, \hat{q}_2| \rightarrow~ _{\textrm{out}}\langle\hat{p}_1', \hat{p}_2'|}\eeq
 and due to the identity
 \beq{_{\textrm{out}}\langle\Omega|a_0^{\textrm{out}}(q_1)a_0^{\textrm{out}}(q_2) = ~_{\textrm{out}}\langle\Omega|a_0^{\textrm{out}}(q_2)a_0^{\textrm{out}}(q_1)}\eeq
 we end up with
 \begin{multline}\tilde{G_{\theta}}^{(2)}(p_1', p_2', \cdots , x_N', x_1, \cdots , x_M) = \frac{\sqrt{Z}}{p_1'^2 - m^2 - i\epsilon}\frac{\sqrt{Z}}{p_2'^2 - m^2 - i\epsilon} e^{-\frac{i}{2}\hat{p}_2'\wedge\hat{p}_1'} \times \\  _{\textrm{out}}\langle\hat{p}_1' \hat{p}_2'|T\left(\phi_{\theta}(x_3')\cdots\phi_{\theta}(x_N')\phi_{\theta}(x_1)\cdots\phi_{\theta}(x_M)\right)|\Omega\rangle + \textrm{OT}.\end{multline}
 The phase can be absorbed to get the twisted out-state
 \beq{\langle\Omega|a_{\theta}^{\textrm{out}}(\hat{p}_2')a_{\theta}^{\textrm{out}}(\hat{p}_1').}\eeq Thus the two-particle residue gives the answer appropriate for the one obtained in Eq.(\ref{shortest}).

 This can be easily generalized to $N$ outgoing particles. For this purpose it, is enough to prove that the phases associated with the outgoing fields give the appropriate phases. This phase comes from manipulating
 \beq\label{25.1}{\langle\Omega|a_{\theta}^{\textrm{out}}(\hat{p}_1')a_{\theta}^{\textrm{out}}(\hat{p}_2')\cdots a_{\theta}^{\textrm{out}}(\hat{p}_N')|\hat{q}_1\cdots\hat{q}_N\rangle}\eeq
 and
 \beq\label{26.1}{\langle \hat{q}_1\cdots\hat{q}_N|a_{\theta}^{\dag}(\hat{p}_N')\cdots a_{\theta}^{\dag}(\hat{p}_1') |\Omega\rangle.}\eeq
 They have phases related by a complex conjugation.  They can be calculated by moving the twists of $a_{\theta}(\hat{p}')$ to the left in Eq.(\ref{25.1}) and to the right in Eq.(\ref{26.1}). This will give the appropriate phase as seen in Eq.(\ref{shortest}).

 We can proceed in a similar manner for incoming particles as well where the conjugates of Eq.(\ref{25.1}) and Eq.(\ref{26.1}) appear. Putting all this together, the final answer is easily seen to be the same as the one obtained in Eq.(\ref{shortest}).


\section{Remarks}

 We have shown two nonperturbative methods relating the commutative and noncommutative scattering amplitudes. Our important result is that
there is complete consistency between the LSZ and interaction representation S-matrix elements on the Moyal plane.

 In the LSZ formulation, the Drinfel'd twist uses the total four-momentum $P_{\mu}$ including interactions. In the interaction representation approach, the Drinfel'd twist instead uses the non-interacting four-momentum $P_{\mu}^0$. No systematic derivation of the interaction representation from an exact formulation using the total four-momentum $P_{\mu}$ is known in the twisted case. The difficulty in this derivation is the appearance of $P_{\mu}$ in the exponential of the Drinfel'd twist. In the absence of this derivation, what is done in the ``interaction representation'' is to use the Drinfel'd twist with the non-interacting four-momentum $P_{\mu}^0$. The equivalence of LSZ and the latter formalism for scattering amplitudes is thus a non-trivial justification of the latter approach. We emphasize that {\it this equivalence does not extend to off-shell Green's functions}~\cite{Future}, a fact which highlights the nontrivial nature of the equivalence of the LSZ and interaction representation scattering theories on the Moyal plane.

The $\theta$-dependence of the S-matrix elements comes through a phase through the products of external momenta. From this it follows immediately that at least for S-matrix elements, there can be no $\theta$-dependence in loop diagrams and hence no possibility of UV-IR mixing in this formalism of quantum scalar fields on the Moyal plane.  Moreover the renormalization of this theory for the S-matrix elements is similar to the one followed for renormalizing the S-matrix elements in the corresponding commutative theory. This in principle completes the renormalization program for S-matrix elements for this scalar noncommutative field theory on the Moyal plane.

The methods followed here can perhaps be extended to include gauge fields as well. Noncommutative field theories with gauge fields involve using a centrally extended Hopf algebra and a corresponding nonassociative spacetime~\cite{BalBabarGauge1}. We leave the computation of scattering amplitudes in such theories to a future paper.

Thermal field theories on the Moyal plane have also been developed through the formalism of thermofield dynamics~\cite{BalTrgThermal, Amilcar}. Scattering amplitudes in noncommutative thermal field theories can be computed in a manner similar to the one shown in this paper~\cite{BalTrgThermal}.

The noncommutative phase, though seen in the scattering amplitude vanishes once the square of the scattering amplitude is taken if the in- and out- states are energy and momentum eigenstates. Such a result is not true once we look at the off-shell $n$-point Green's functions of twisted quantum fields where the $\theta$-dependence is not through an overall phase. Such a $\theta$-dependence can have consequences for the $\beta$-function of this theory. This will be reported in a forthcoming paper.

\section{Acknowledgements}
It is a pleasure to thank Prof. Sachin Vaidya, Prof. T.R.Govindarajan and Rahul
Srivastava for many useful discussions and critical comments on the notes prepared for this
paper. They also thank the Institute of Mathematical Sciences, Chennai and especially 
Prof. T.R.Govindarajan for warm hospitality and support during the course of this work.
Thanks are also due to Prof. Alvaro Ferraz for his kind hospitality at IIP-UFRN, Natal when this work was being written up.

 APB and PP were supported by DOE under the grant number
DE-FG02-85ER40231. ARQ is supported by  CNPq under grant number 307760/2009-0.

\end{document}